\documentclass[a4paper]{article}
\usepackage{array}
\usepackage{INTERSPEECH2021}
\usepackage{graphicx}
\usepackage{caption}
\usepackage{subcaption}
\usepackage{multirow}
\usepackage{tabularx}
\usepackage{float}
\usepackage{textgreek}
\title{Efficient Attention Branch Network with Combined Loss Function for Automatic Speaker Verification Spoof Detection}
\name{Amir Mohammad Rostami$^1$, Mohammad Mehdi Homayounpour$^1$, Ahmad Nickabadi $^1$}
\address{
  $^1$Department of Computer Engineering, Amirkabir University of Technology}
\email{[a.m.rostami, homayoun, nickabadi]@aut.ac.ir}

\begin{document}

\maketitle
\begin{abstract}
  Many endeavors have sought to develop countermeasure techniques as enhancements on Automatic Speaker Verification (ASV) systems, in order to make them more robust against spoof attacks. As evidenced by the latest ASVspoof 2019 countermeasure challenge, models currently deployed for the task of ASV are, at their best, devoid of suitable degrees of generalization to unseen attacks. Upon further investigation of the proposed methods, it appears that a broader three-tiered view of the proposed systems; comprised of the classifier, feature extraction phase, and model loss function, may to some extent lessen the problem. Accordingly, the present study proposes the Efficient Attention Branch Network (EABN) modular architecture with a combined loss function to address the generalization problem. The EABN architecture is based on attention and perception branches; the purpose of the attention branch—also interpretable from a human’s point of view—is to produce an attention mask meant to improve classification performance. The perception branch, on the other hand, is used for the primary purpose of the problem at hand, that is, spoof detection. The new EfficientNet-A0 architecture was employed for the perception branch, with nearly ten times fewer parameters and approximately seven times fewer floating-point operations than the top performing SE-Res2Net50 network. The final evaluation results on ASVspoof 2019 dataset suggest an EER = 0.86\% and t-DCF = 0.0239 in the Physical Access (PA) scenario using the logPowSpec input feature, the EfficientNet-A0 for the perception branch, and the combined loss function. Furthermore, using the LFCC input feature, the SE-Res2Net50 for the perception branch, and the combined loss function, the proposed model performed at figures of EER = 1.89\% and t-DCF = 0.507 in the Logical Access (LA) scenario, which to the best of our knowledge, is the best single system ASV spoofing countermeasure. 
\end{abstract}

\section{Introduction}
Remote authentication has excited great interests in various academic circles and otherwise, given the increasing reliance on online applications as well as the onset of certain conditions such as the Covid-19 pandemic. Such circumstances call for an easy-to-use, accurate, and efficient authentication system. Along this thread, Automatic Speaker Verification (ASV) system and other biometric systems such as face recognition, electronic signatures, iris-based, and hybrid methods have been proposed as a means to satisfy user needs \cite{1reynolds1995robust,2hansen2015speaker}. Nevertheless, virtually all of these systems are vulnerable to spoof attacks (i.e., spoofable). A system based on face recognition for example, may be spoofed by simply displaying a person's image (photo) to the system \cite{3hadid2015biometrics}. Likewise, a fingerprint system can be spoofed by copying a fingerprint. In particular, ASV systems are also vulnerable in the face of four types of attacks, counting on recording and replaying the voice of the authorized person (replay attack), text-to-speech systems that are trained with the voice of the targeted person, voice conversion systems, and speaker imitation \cite{4ergunay2015vulnerability}. The threats facing ASV systems in terms of spoof attacks are potentially high and may amass to serious implications \cite{5sahidullah2019introduction}. In consequence, since 2015, ASVspoof challenges were held for research communities worldwide to try and enhance ASV systems so as to make them robust against spoofing attacks. 

A total of three ASVspoof challenges have thus far been held, with the first instance in 2015, covering only speech synthesis and voice conversion (also called logical access scenario) attacks \cite{6wu2015asvspoof}. A variety of methods and systems were proposed and implemented by ASV organizers to produce spoof samples, exciting the interest of many researchers intrigued by both the challenge and the dataset provided therein. The second ASVspoof challenge held in 2017, focused more on the replay attack (also called physical access scenario) \cite{7delgado2018asvspoof,8kinnunen2017asvspoof}. In order to be able to test the performance of countermeasure systems in real conditions, the organizers produced the dataset in different environmental conditions and using different devices. Further comprehensive conditions were investigated in 2019 to account for all three attacks considered in previous challenges \cite{9todisco2019asvspoof}, ushering in the development of an extensive dataset using state-of-the-art voice conversion and speech synthesis systems. Spoofing samples in this challenge were more realistic and challenging in view of the improvements made to the spoof systems in previous years. For replay attacks, in particular, samples were produced with greater degrees of control, and a tandem detection cost function (t-DCF) metric was used as the primary metric to assess the efficiency of integrating countermeasures with ASV systems.

Models proposed to assess the ASVspoof 2019 dataset can be categorized into two main classes: methods based on extraction and engineering of features and methods based on classifier architecture. Methods of the first category incorporate features such as Mel-filter frequency Cepstral coefficients (MFCC), Inverted Mel-Filter Frequency Cepstral coefficients (IMFFC), Constant Q Cepstral Coefficients (CQCC), Group Delay (GD) gram, Instantaneous Amplitude (IA), Instantaneous Frequency (IF), X-vectors, and features from deep learning models \cite{10cai2019dku,11todisco2016new,12alam2018boosting,13suthokumar2018modulation,14li2018multiple,15xiao2015spoofing,27kamble2020amplitude,34qian2016deep}. Some methods also use raw signals to extract features using methods such as SincNet \cite{16zeinali2019detecting} or Variational Auto Encoder (VAE) \cite{17chettri2020deep}. The second category deals with a variety of classifiers such as Neural network-based methods including VGG \cite{16zeinali2019detecting}, Squeeze-Excitation (SE), Residual network, Siamese networks \cite{18lai2019assert}, and recurrent networks \cite{19huang2020audio}, as well as other traditional GMM-based methods. Certain methods have also used end-to-end structures for this purpose.

Inquiries made into the 2019 ASVspoof dataset results are suggestive of two primary drawbacks of the proposed methods. The first points to a lack of generalization and high error rate against unseen attacks, which is clearly observed given the difference between errors obtained for the training, development, and evaluation sets. In addressing this lack of generalization, numerous studies have tried to improve generalization by means of fusing several models (ensemble models) \cite{20korshunov2017impact}. Such fusion and ensemble models and methods that use deep neural networks have led to considerable increases in model parameters as well as the necessary floating-point operations (FLOPS). Under such circumstances, it would be infeasible to use the proposed models in specific applications. This provides the required grounds for the integration of simple yet efficient countermeasure techniques with ASV systems to make ASV robust. Moreover, the proposed body of research fails to provide a detailed understanding for how models detect spoof attacks or handle generalization issue. This ambiguity can be interpreted in terms of the incapacity of humans or rather human-oriented decision making to differentiate between the spoofed and the bonafide samples detected by the final system. A detailed examination of this issue can provide further insight into the development of better system.

The primary purpose of this work is to provide a model for detecting spoofing attacks on ASV systems. An interpretable attention mask in a new modular architecture was used for this purpose via the introduction of perception and attention branches. Furthermore, for the first time in this domain, the EfficientNet-A0 \cite{21rostami2020efficientnet} architecture was employed to achieve a system with low number of parameters and FLOPS. The proposed architecture along with the newly combined loss function and masks that provide a more human-oriented perspective, was used to obtain comparable and, in some cases, top-performing results in these spoofing attacks. 

The following section provides a brief review of relevant studies conducted in recent years. The proposed countermeasures and the loss function are introduced in Section 3 and Section 4 calls attention to the general configuration used for experiments. Section 5 gives the analysis results along with a summarization of the work. The study is finally concluded in Section 6.
\section{Related work}

This section reviews some of the research carried out on spoof attack detection, taking a look on the best-performing methods, as per results obtained on the ASVspoof 2019 dataset. Similar models and tasks were also investigated inclusive of new architecture and the application of attention mechanisms and attitude in the loss function.

Cheng-I Lai et al. proposed a deep model to obtain discriminative features in both time and frequency domains \cite{22lai2019attentive}. The proposed design includes a filter-based attention mechanism used to improve or ignore commonly extracted features implemented in the ResNet architecture to classify attended input maps. The reserved classifier used in their study (Residual Network) consists of a convolution layer equipped with dilated mechanism instead of a fully connected layer, which runs as an attentive filtering network; i.e., masks input features. The obtained results were suggestive of the relatively high performance of the model given the use of an attention mechanism to produce attention masks as well as an appropriate classifier.

X. Li et al. attempted to use the Res2Net architecture, which has achieved significant results in various computer vision tasks \cite{23li2021replay}. They proposed a new Res2Net architecture by revisioning ResNet blocks to allow for multi-scale features. In a Res2Net architecture, input feature maps of a block are divided into several groups of channels with a similar residual structure to the original ResNet. Using channels, feature map sizes can be different, increasing the covered area, and thereby yielding features with different scales. This modification improves system performance and the model's generalization against unseen attacks. In addition, using this architecture could reduce the size or number of model parameters relative to the original ResNet structure while improving model performance. The obtained results show that the Res2Net50 model performs better than the ResNet34 and ResNet50 models in both physical and logical access scenarios. They also showed that integrating the block with Squeeze-and-excitation (SE), which produces SE-Res2Net blocks, leads to better performance. Figure 1 illustrates the architecture and structure of these blocks. Significant results were also obtained in both scenarios for the proposed SE-Res2Net50 network based on SE-Res2Net blocks and CQT feature. The network proposed in this work has nearly 0.9 million parameters, which is relatively small compared to other architectures. However, the main drawback to the model is the high number of FLOPS, which leads to increased runtime in the inference phase due to the multiplicity of blocks and the structure of SE-Res2Net.
\begin{figure}[H]
    \centering
    \begin{subfigure}[b]{0.12\textwidth}
        \centering
        \includegraphics[width=\textwidth]{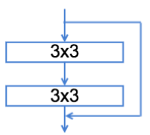}
        \caption{ResNet}
    \end{subfigure}
    \hfill
    \begin{subfigure}[b]{0.12\textwidth}
        \centering
        \includegraphics[width=\textwidth]{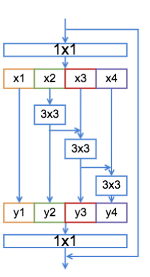}
        \caption{Res2Net}
    \end{subfigure}
    \hfill
    \begin{subfigure}[b]{0.12\textwidth}
        \centering
        \includegraphics[width=\textwidth]{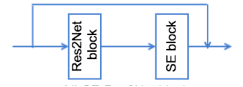}
        \caption{SE-Res2Net}
    \end{subfigure}
    \hfill
    \label{fig:seres}
    \caption{ResNet, Res2Net, and SE-Res2Net blocks \cite{23li2021replay}.}
\end{figure}
Zhang et al. focused on logical attacks in their work \cite{24zhang2021one}, explaining the lack of model generalization against unseen attacks as caused by the formulation of the spoof detection problem as a binary classification. The difficulty with using a binary classifier can be interpreted in terms of the distribution of training and test data for spoof and bonafide samples as not being the same. More specifically, samples in the test set generated by new systems or conditions not found in training data cause differences in distribution; which, however, is not the case for bonafide samples. The problem was, therefore, redefined as a one-class classification problem, where the distribution of a target class for a specific problem should be the same in both training and test datasets, irrespective of whether other classes have similar distributions or not. In such cases, the primary objective is to obtain the bonafide distribution and define a rigid decision boundary around it so that unseen samples from other classes cannot cross that decision boundary. To this aim, a one-class softmax loss function was incorporated for learning a feature space that can map bonafide samples in a dense space, while maintaining a good margin with spoofing samples. Finally, by means of the ResNet-18 network and the LFCC features, the authors succeeded in attaining top-performing results for logical access attacks.
\begin{figure}[H]
  \centering
  \includegraphics[width=\linewidth]{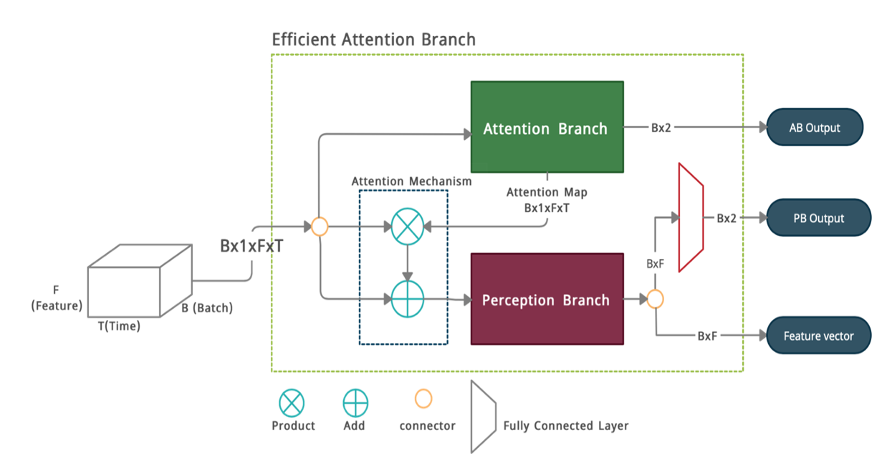}
  \caption{Proposed Efficient Attention Branch Network architecture.}
  \label{fig:AttentionBranchNetwork}
\end{figure}
\section{Proposed Model}
\subsection{Network Architecture}
The overall architecture of the proposed network was designed with three main objectives in mind; that the architecture be small enough to explicate an appropriate number of parameters, while maintaining an acceptable runtime in order to achieve satisfactory performance in most ASV applications; the architecture was designed to be interpretable by humans. To put differently, the architecture was required to somehow express what discriminates bonafide speech from speech made in a spoof attack as a means to improve systems in the future; lastly, the model was configured to emulate comparable performance to relevant classifiers used for this purpose.

To achieve all these goals, the Efficient Attention Branch Network (EABN) was proposed in this study. The intended framework adopts a well-performed Attention Branch Network \cite{25fukui2019attention} in computer vision as the main idea for the EABN architecture. As shown in Figure \ref{fig:AttentionBranchNetwork}, this network consists of two branches of attention and perception. The attention branch seeks to improve the performance of the perception branch by means of producing an attention mask, which is then applied to make the discriminative parts of the input feature map more. In addition to improving the performance of the perception branch, masks produced by the attention branch are also interpretable from a human point of view; as these masks are used for dummy classification tasks. The primary work load is performed in the perceptual branch, where the probability output of each class is produced.
\subsection{Attention Branch}
\begin{figure}[H]
  \centering
  \includegraphics[width=0.9\linewidth]{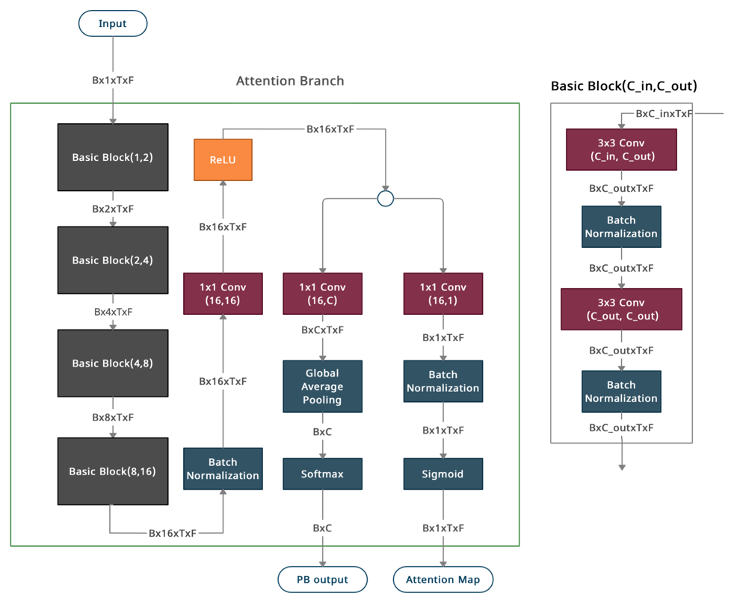}
  \caption{Proposed architecture for Attention branch.}
  \label{fig:AttentionBranch}
\end{figure}

The attention branch itself comprises of two main parts, as shown in Figure \ref{fig:AttentionBranch}. As can be observed, the input feature map is initially fed into the attention branch, which uses four consecutive Basic Blocks to extract the appropriate features and to convert the input features to 16-feature maps. The blocks consist of two convolution layers with 3×3 kernels, which are then linked to the batch normalization layer. In addition to feature extraction, the first convolution layer also increases the feature map size, while the second convolution layer exclusively handles the feature extraction process. The obtained feature maps are eventually transformed from a 16-size map to a single feature using a convolution layer with a 1×1 kernel, which then goes through a softmax layer to yield the final output attention mask.

The other branch produces a human-interpretable attention mask. This is carried out by using a convolution operation to transform the 16 feature maps into maps the same size as the number of classes for the problem—which in this study includes the bonafide and spoof classes. Then, using a Global average Pooling layer, these two feature maps are converted to a 2×1  tensor. Finally, by applying softmax, the probability of a feature map belonging to each class is obtained. These probabilities are later used in the optimization process for the proposed combined loss function. Through the process of optimization, feature maps are generated so that in addition to helping the perception branch, they can also be used for classification and be made interpretable from a human perspective.
\subsection{Perception Branch}
\begin{figure}[t]
  \centering
  \includegraphics[width=0.8\linewidth]{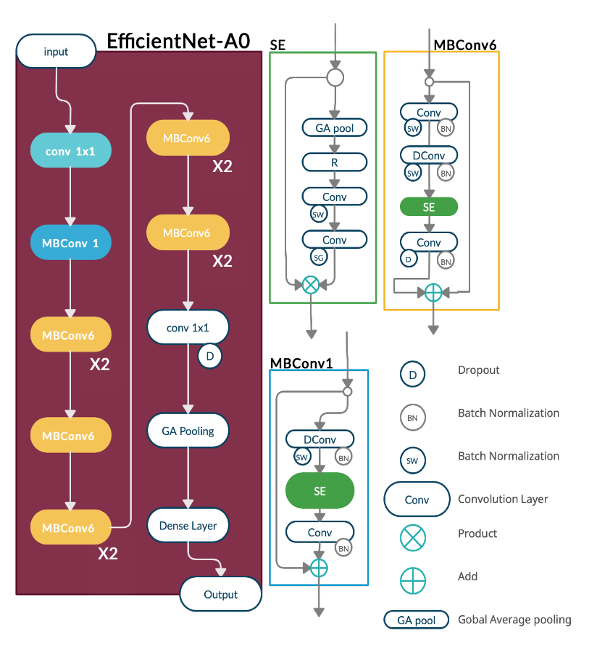}
  \caption{Proposed architecture for perception branch operating via the reverse compound scaling method.}
  \label{fig:PerceptionBranch}
\end{figure}

The perception branch can constitute virtually any classifier. However, as the primary objectives of this study call for low number of parameters, low runtime, and good performance in network design, the EfficientNet architecture—commonly introduced as a high performing model in image classification tasks and speech processing tasks such as speech recognition and keyword spotting—was employed. The fundamental architecture of the EfficientNet family is called EfficientNet-B0, which has about 4 million parameters. This number of parameters is not suitable for the target applications of this study. Alternatively, the approach introduced in the EfficientNet-Absolute Zero work \cite{21rostami2020efficientnet}, which applies the reverse of the compound scaling method, was used. The scaling method ($S$) is designed to shrink a base model ($M$) by decreasing the depth ($\alpha$), width ($\beta$), and resolution of the input image ($\gamma$), simultaneously. A formulation of this method is given below as an optimization problem, in which the goal is to satisfy the intended conditions so that the final model has the best performance.
$$
\begin{aligned}
&\begin{aligned}
max_{d,r,w}\ Accuracy(S(M,\ d,\ r,\ w))
\end{aligned}\\
&\qquad\text { s.t. }\\
&\qquad\qquad\frac{1}{20} \leq \alpha \cdot \beta^{2} \cdot \gamma^{2} \leq \frac{1}{16}\\
\qquad\qquad&\begin{aligned}
\qquad\qquad&0.2 \leq \alpha,\ \beta \leq 0.6,\ \gamma=2
 \\
\end{aligned}
\end{aligned}
$$
The two parameters α and β are set by applying a grid search on intervals [$0.2-0.6]$ with steps of 0.005. Eventually, 19 models were evaluated with a small subset of samples, with parameters $\alpha$ and $\beta$ set at values 0.2 and 0.25, respectively. $\gamma$ was also set at $\approx$ 2, given the input image size (513×400) and EfficieNet-B0 input-size of 256×256. Figure \ref{fig:PerceptionBranch} illustrates the final model obtained for the perception branch with 95,000 parameters.
The input to this branch is $m(x_i)$, where $x_i$ is the input image for the $i^{th}$ sample and is calculated from the following equation:
$$
m\left(x_{i}\right)=\left(1+g\left(x_{i}\right)\right) \times x_{i}
$$
where $g(x_i)$ is the attention mask produced for the $i^{th}$ sample by attention branch. The output of this network is a vector of length 256, which represents the embedded vector of the input image and is applied for two scenarios: the first uses the vector along with a fully connected layer and the softmax layer to yield probabilities for each individual sample; the second scenario uses the vector as input to a loss function. Thus, samples are embedded in a 256-dimensional space in the most distinctive way.
\subsection{Loss function}
To train model parameters, a combined loss function was used to account for all study objectives. To train an attention branch capable of producing interpretable masks, the $AB_{output}$ was used as input to a weighted Cross-Entropy (CE) loss function. It should be noted that by introducing the proposed loss function with coefficient $\lambda_{AB}$, values in equation \ref{eq:3} are altered. Proceeding forward, the Triplet Center Loss (TCL) function was used to train the embedding vectors. TCL works in the same way as the triplet loss function, except that it no longer needs to mine triplets for training, and this difference makes the training process faster and more stable. This loss function considers center points for each class in the problem, which are initially assigned random values. The loss function then approaches to make it so that samples of one class are close to the center of their class and away from the nearest center of other classes. In other words, each sample tries to be closer to the center of its class and away from the remaining centers.  The two centers used in this study to represent spoof and bonafide samples were $C_{spoof}$ and $C_{bonafide}$, respectively. The goal here was to ensure that bonafide samples be close to the center of their respective target class, $C_{bonafide}$, and away from $C_{spoof}$. As a result, samples of a specific class in a dense space are closer to each other; representing feature vectors embedded for each sample in the desired space. TCL can be obtained for the $x_i$ sample as follows:
\begin{equation}
L_{\text {focal }}\left(p_{t}\right)=-\alpha_{t}\left(1-p_{t}\right)^{0.005} \log \left(p_{t}\right)
\end{equation}
\begin{equation}
L_{P B}\left(x_{i}\right)=L_{t c}\left(x_{i}\right)+\lambda_{\text {focal }} L_{\text {focal }}\left(x_{i}\right)
\end{equation}
\begin{equation}\label{eq:3}
L_{\text {total}}=L_{PB}+\lambda_{AB} L_{\mathrm{AB}}
\end{equation}
\begin{table}[H]
\caption{Summary of the ASVspoof2019 dataset.}
\centering
\begin{tabular}{|c|cc|cc|}
\hline
\multirow{2}{*}{\textbf{Partition}} & \multicolumn{2}{c|}{\textbf{PA}} & \multicolumn{2}{c|}{\textbf{LA}} \\ 
   & \# Spoof& \# Bonafide& \# Spoof& \# Bonafide \\ \hline
Train & \multicolumn{1}{c|}{48600}& \multicolumn{1}{c|}{5400}& \multicolumn{1}{c|}{22800}& \multicolumn{1}{c|}{2580}\\\hline
Dev & \multicolumn{1}{c|}{24300}& \multicolumn{1}{c|}{5400}& \multicolumn{1}{c|}{22296}& \multicolumn{1}{c|}{2548}\\\hline
Eval & \multicolumn{1}{c|}{116640}& \multicolumn{1}{c|}{18090}& \multicolumn{1}{c|}{63882}& \multicolumn{1}{c|}{7355}\\\hline
\end{tabular}
\label{tab:dataset}
\end{table}
\begin{table*}[th]
\centering
\caption{t-DCF hyperparameters value.}
\begin{tabular}{|c|ccc|cc|cc|}
\hline
\multirow{2}{*}{\textbf{Attack Type}} & \multicolumn{3}{c|}{\textbf{Probabilities}} & \multicolumn{2}{c|}{\textbf{ASV costs}} & \multicolumn{2}{c|}{\textbf{Countermeasure costs}} \\ 
   & $\pi_{\mathrm{tar}}$                           & $\pi_{\mathrm{non}}$                           & $\pi_{\mathrm{spoof}}$    & $C_{\mathrm{fa}}^{\mathrm{asv}}$  & $C_{\mathrm{miss}}^{\mathrm{asv}}$ & $C_{\mathrm{fa}}^{\mathrm{cm}}$  & $C_{\mathrm{miss}}^{\mathrm{cm}}$ \\ \hline
PA & \multicolumn{1}{c|}{0.9405} & \multicolumn{1}{c|}{0.0095} & 0.05 & 10 & 1 & 10 & 1 \\ \hline
LA & \multicolumn{1}{c|}{0.9405} & \multicolumn{1}{c|}{0.0095} & 0.05 & 10 & 1 & 10 & 1 \\ \hline
\end{tabular}
\label{tab:datasethyperparam}
\end{table*}
\section{Experimental configuration}
\subsection{Dataset and evaluation metrics}
The proposed method was evaluated using the ASVspoof 2019 dataset, which includes two scenarios: physical access (PA) and logical access (LA). Details of this dataset are shown in Table \ref{tab:dataset}. Furthermore, considering that one of the objectives of this research is the simultaneous use of countermeasure and ASV system, the tandem-detection cost function (t-DCF) and the equal error rate (EER) metrics were used. This metric was introduced as the primary evaluation metric of the 2019 challenge, which is calculated as:
$$
\mathrm{t}-\mathrm{DCF}(s)=C_{1} P_{\mathrm{miss}}^{\mathrm{cm}}(s)+C_{2} P_{\mathrm{fa}}^{\mathrm{cm}}(s)
$$
where
$P_{\mathrm{fa}}^{\mathrm{cm}}(s)$ 
and
$P_{\mathrm{miss}}^{\mathrm{cm}}$ 
are the false acceptance error rate and the false rejection error rate of the countermeasure, respectively. Considering the threshold, s, values for the two error rates can be obtained as follows:
$$
\begin{aligned}
P_{\text {miss }}^{\mathrm{cm}}(s) &=\frac{\#\{\text { bona fide trials with CM score } \leq s\}}{\#\{\text { Total bona fide trials }\}} \\
P_{\mathrm{fa}}^{\mathrm{cm}}(s) &=\frac{\#\{\text { spoof trials with CM score }>s\}}{\#\{\text { Total spoof trials }\}}
\end{aligned}
$$
The two constants $C_1$ and $C_2$ represent the predefined cost for the errors, which are determined based on prior probabilities as shown below:
$$
\left\{\begin{array}{l}
C_{1}=\pi_{\operatorname{tar}}\left(C_{\mathrm{miss}}^{\mathrm{cm}}-C_{\mathrm{miss}}^{\mathrm{asv}} P_{\mathrm{miss}}^{\mathrm{asv}}\right)-\pi_{\mathrm{non}} C_{\mathrm{fa}}^{\mathrm{asv}} P_{\mathrm{fa}}^{\mathrm{asv}} \\
C_{2}=C_{\mathrm{fa}}^{\mathrm{cm}} \pi_{\mathrm{spoof}}\left(1-P_{\mathrm{miss}, \mathrm{spoof}}^{\mathrm{asv}}\right)
\end{array}\right.
$$
Here, $C_{\mathrm{miss}}^{\mathrm{asv}}$ represents the cost incurred by the error of the ASV system for the false rejection error rate of the genuine person, and $C_{\mathrm{fa}}^{\mathrm{asv}}$ represents the false acceptance error rate when ASV authorizes the wrong person. Each countermeasure error also corresponds to two costs; $C_{\mathrm{miss}}^{\mathrm{cm}}$, which indicates the cost in recognizing a bonafide sample as a spoof, and $C_{\mathrm{fa}}^{\mathrm{cm}}$, which indicates a mistake in accepting a sample produced by a spoof system as bonafide. In addition, the probability of occurrence of any class of genuine ($\pi_{\mathrm{tar}}$), non-target or imposter ($\pi_{\mathrm{non}}$) and spoof attack ($\pi_{\mathrm{spoof}}$) are also considered with the condition $\pi_{\mathrm{tar}}+\pi_{\mathrm{non}}+\pi_{\mathrm{spoof}}=1$. Cost and probability values are calculated as in Table \ref{tab:datasethyperparam}.
\subsection{Feature extraction and engineering}
Based on past researches and works, a single acoustic feature was considered for each of the attacks. For the PA scenario, we used the logarithm power of the spectrogram (logPowSpec) with 25 ms frames, 10ms step size with 1024 samples (with zero padding applied if needed), using Hamming window. All the samples are first transformed into 4s voice segments. To do this, samples that are less than 4 seconds are repeated to achieve a 4s segment. Longer samples are also divided into 4s segments with no overlap, and each segment is considered an individual utterance. The final input form consists of a spectrogram with 513×400 dimensions. For the LA scenario, the LFCC feature was extracted according to the procedure used in the base model presented in the ASVspoof 2019 challenge. Here, 20ms frames with 512 Fourier transform points and 20 filters were used along with their first and second derivatives. Finally, a two-dimensional tensor with dimensions of 60×400 was obtained. 

As a further step, specAug \cite{26park2019specaugment} techniques were applied for better training and generalization. This method has worked well for other speech tasks, such as speaker verification, speech recognition, and keyword potting \cite{21rostami2020efficientnet}. The method was implemented by applying zero masks on the time and frequency axis for each training sample with a probability of 0.25. The size of this band is randomly between 20 and 80 frames on the time axis. In the frequency axis; the size of this band is selected randomly between 5 and 20 for the LFCC and between 25 and 100 for the logPowSpec. As a result, the model is capable of accounting for all time frames and frequency bins and preventing overfitting at specific time or frequency points.
\subsection{Perception branch models}
In addition to the proposed EfficientNet-A0 architecture, a SE-ResNet50 architecture was also used, which achieved significant results. The models were then compared in terms of both efficiency and performance, and the EABN modularity idea was evaluated accordingly.
\subsection{Training procedure}
The final results obtained from experiments on small subsets of the ASVspoof 2019 dataset yielded values of 0.1, 0.005, and 32 for $\lambda_{AB}$, $\lambda_{focal}$ and $m$, respectively. To optimize the loss function with assigned values, configurations for the SE-ResNet50 architecture were adopted. In the case of Adam optimization, $\beta_1$, $\beta_2$, and learning rate were obtained at $0.9$, $0.98$, and $10^{-9}$, respectively. The learning rate initially drops linearly for the first 1000 steps and then decreases in proportion to the inverse of the square root of the number of steps. All models were trained with 40 epochs and the model with the lowest EER on the development set of the dataset was selected as the optimal choice. Batch-sizes were set at 64 and 128 when using EfficienNet-A0 as the perception branch module with LFCC and logPowSpec, respectively. Due to the relatively greater number of parameters for the SE-Res2Net50 model compared to EfficientNet-A0, a batch-size of 8 was used for LFCC and logPowSpec features. The models were implemented on a GTX-1080ti GPU on Linux OS. The source code of our implementations based on Python and Pytorch is publicly available\footnote{https://github.com/AmirmohammadRostami/ASV-anti-spoofing-with-EABN}.
\section{Results}
\subsection{Perception branch’s models evaluation}
This section evaluates the overall architectural EABN and the EfficientNet-A0 network as a classifier for spoof detection. To investigate EABN performance, the EfficientNet-A0 and SE-ResNet50 architectures were used for the perception branch, which have the lowest EER as a single model to the best of our knowled.ge. The results for both attacks are shown in Table 3. In the PA scenario, EfficientNet-A0 showed better performance than SE-ResNet50 and nearly ten times fewer parameters and seven times fewer FLOPS. On the other hand, the SE-ResNet50 model performs better when using LFCC feature for the LA scenario. This can be explained in terms of the enhanced performance of the EfficientNet-A0 model in extracting features from the spectrogram. On the other hand, the SE-ResNet50 model works best when the feature input is processed, and the EfficientNet-A0 model converges very quickly or suffers from overfitting.
\begin{table*}[]
\caption{Result of models used in perception branch and input features on ASVspoof 2019 evaluation dataset for PA and LA scenarios. K, M, and G represent Kilo, Mega, and Giga, respectively. }
\centering
\begin{tabular}{|c|c|c|c|c|c|c|c|c|}
\hline
\multirow{2}{*}{\textbf{\#}} & \multirow{2}{*}{\textbf{Perception branch model}} & \multirow{2}{*}{\textbf{Input feature}} & \multirow{2}{*}{\textbf{\#Parameters}} & \multirow{2}{*}{\textbf{\#Flops}} & \multicolumn{2}{c|}{\textbf{PA}}  & \multicolumn{2}{c|}{\textbf{LA}}  \\ \cline{6-9} 
                             &                                                   &                                         &                                        &                                   & \textbf{EER(\%)} & \textbf{t-DCF} & \textbf{EER(\%)} & \textbf{t-DCF} \\ \hline
1                            & EfficientNet-A0                                   & LFCC                                    & \multirow{2}{*}{95k}                   & 198M                              & -                & -              & 3.68             & 0.0931         \\ \cline{1-3} \cline{5-9} 
2                            & EfficientNet-A0                                   & LogPowSepc                              &                                        & 1.696G                            & 0.86             & 0.0239         & -                & -              \\ \hline
3                            & SE-Res2Net50                                      & LFCC                                    & \multirow{2}{*}{964k}                  & 1.519G                            & -                & -              & 1.89             & 0.0597         \\ \cline{1-3} \cline{5-9} 
4                            & SE-Res2Net50                                      & LogPowSepc                              &                                        & 12.929G                           & 0.98             & 0.2769         & -                & -              \\ \hline
\end{tabular}
\label{tab:PerModelComp}
\end{table*}
\begin{figure}[H]
    \centering
    \begin{subfigure}[b]{0.22\textwidth}
        \centering
        \includegraphics[width=\textwidth]{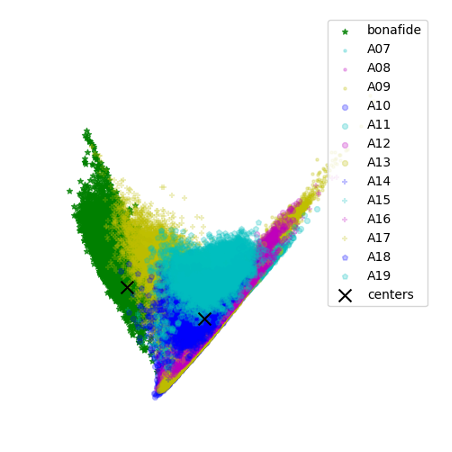}
        \caption{}
    \end{subfigure}
    \hfill
    \begin{subfigure}[b]{0.22\textwidth}
        \centering
        \includegraphics[width=\textwidth]{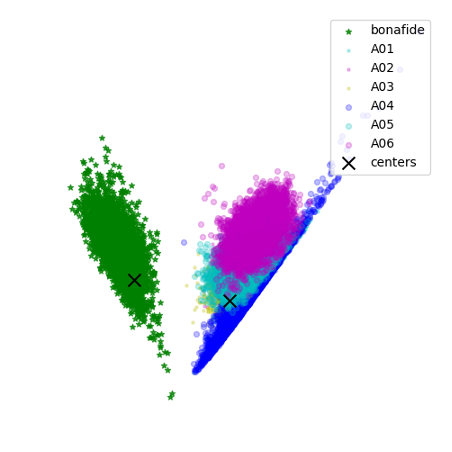}
        \caption{}
    \end{subfigure}
    \hfill
    \caption{Feature embedding visualization of our proposed loss function for evaluation (a) and training (b) sets of the ASVspoof 2019 LA attack. Features were reduced to 2-D space using PCA.}
    \label{fig:lossfunction}
\end{figure}
\subsection{Loss function}
The proposed combined loss function was used for the first time in this work to achieve a discriminative vector space to distinguish spoof samples from bonafide samples. More precisely, the triplet center loss was used to map input samples to a discriminative space. As shown in Figure \ref{fig:lossfunction}, the training samples mapping space is suitable for the classification problem. Examining test samples that include unseen attacks also demonstrate that the resulting space is reasonably discriminative. It can therefore be said that the model shows good generalization for unseen attacks. The best value for margin 32 was obtained in this study by testing three values of 16, 32, and 64. 
\subsection{Attention masks}
\begin{figure}[H]
    \centering
    \begin{subfigure}[b]{0.23\textwidth}
        \centering
        \includegraphics[width=\textwidth]{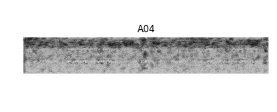}
    \end{subfigure}
    \hfill
    \begin{subfigure}[b]{0.23\textwidth}
        \centering
        \includegraphics[width=\textwidth]{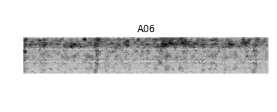}
    \end{subfigure}
    \hfill
    \begin{subfigure}[b]{0.23\textwidth}
        \centering
        \includegraphics[width=\textwidth]{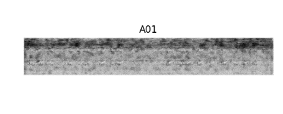}
    \end{subfigure}
    \hfill
    \begin{subfigure}[b]{0.23\textwidth}
        \centering
        \includegraphics[width=\textwidth]{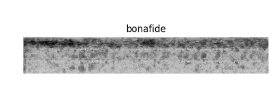}
    \end{subfigure}
    \hfill
    \begin{subfigure}[b]{0.23\textwidth}
        \centering
        \includegraphics[width=\textwidth]{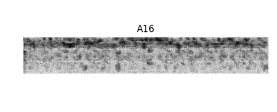}
    \end{subfigure}
    \hfill
    \begin{subfigure}[b]{0.23\textwidth}
        \centering
        \includegraphics[width=\textwidth]{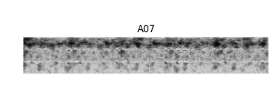}
    \end{subfigure}
    \hfill
    \caption{Average of produced LFCC attention masks for some spoof attacks in ASVspoof 2019 evaluation. As can be seen, the attention branch tries to dominant high frequencies to detect spoof attacks.}
    \label{fig:lfcc}
\end{figure}

\begin{figure}[H]
\centering
    \begin{tabular}{|m{0.75cm}m{1.7cm}m{1.7cm}m{1.7cm}|}
    \hline
    \textbf{Class} &
    \textbf{Inp. feature} &
    \textbf{Att. mask} &
    \textbf{Perc. input}
    \\
    A03 &
    \includegraphics[width=1.6cm]{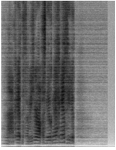} &
    \includegraphics[width=1.6cm]{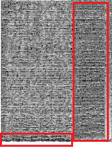} &
    \includegraphics[width=1.6cm]{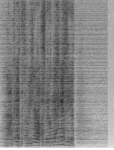}
    \\
    A09 &
    \includegraphics[width=1.6cm]{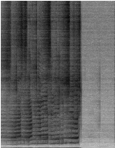} &
    \includegraphics[width=1.6cm]{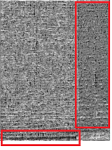} &
    \includegraphics[width=1.6cm]{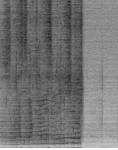}
    \\
    A05 &
    \includegraphics[width=1.6cm]{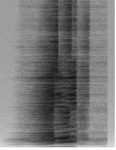} &
    \includegraphics[width=1.6cm]{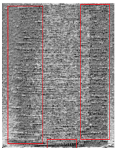} &
    \includegraphics[width=1.6cm]{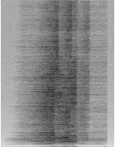}
    \\
    A06 &
    \includegraphics[width=1.6cm]{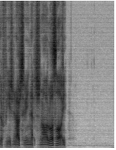} &
    \includegraphics[width=1.6cm]{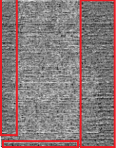} &
    \includegraphics[width=1.6cm]{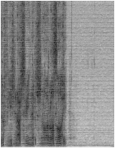}
    \\
    B &
    \includegraphics[width=1.6cm]{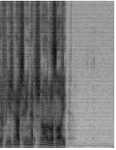} &
    \includegraphics[width=1.6cm]{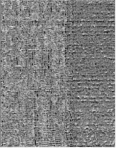} &
    \includegraphics[width=1.6cm]{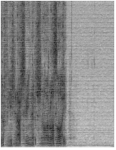}
    \\
    \hline
    \end{tabular}
\caption{Input (Inp.) feature, produced attention (Att.) mask, and final input feature for perception (Perc.) branch of some samples in the evaluation set for logPowSpec feature (B is bonafide class). Red boxes are parts of input features that the attention branch wants to dominate and are interpretable from a human’s perspective.}
\label{fig:spec}
\end{figure}

One of the main concerns about the proposed architecture was to obtain attention masks that could be interpreted from a human point of view. This was investigated for the LFCC feature, with the averaged masks generated for all samples in the evaluation set shown in Figure \ref{fig:lfcc}. As for the logPowSpec mask, few samples from the evaluation set of the PA attack are shown in Figure \ref{fig:spec}. The raw input features and results of the applied mask on the input feature, which is input for the perception branch, are also shown in this figure.
Examining the LFCC feature masks obtained for different attack systems reveals that higher-frequency domain information is more effective in detecting spoof patterns. This is consistent with the fact that models using LFCC features tend to outperform those that work with MFCCs in most studies. As presented by the findings, resolution values are lower at high frequencies for MFCC than LFCC. Examining the masks created for physical access attacks also show that the model emphasizes silent parts of speech. This is in all likelihood to the fact that the effects of record and play devices on recorded speech are more considerable during silences, and the countermeasure can more easily detect attacks accordingly. Convolution operations in the perception branch also tend to converge faster by blurring and dominating some values at different frequencies. This can in effect be achieved by decreasing the impacts of frequencies that show lower capacity to discriminate spoof attacks from bonafide samples. In this regard, it can be said that paying attention to silence intervals and reducing the impact of human speech frequencies lead to better detection of physical access attacks.

\begin{table*}[t]
\caption{Performance comparison of the proposed systems with known single systems tested on the ASVspoof  2019 PA and LA evaluation set. Models are named base on their input feature, the classification model, and the loss function.}
\centering
\begin{tabular}{|c|c|c|c|c|}
\hline
\multirow{2}{*}{\textbf{Input feature + Classifier + Loss function}} &
  \multicolumn{2}{c|}{\textbf{PA}} &
  \multicolumn{2}{c|}{\textbf{LA}} \\ \cline{2-5} 
 &
  \multicolumn{1}{c|}{\textbf{EER(\%)}} &
  \multicolumn{1}{c|}{\textbf{t-DCF}} &
  \multicolumn{1}{c|}{\textbf{EER(\%)}} &
  \multicolumn{1}{c|}{\textbf{t-DCF}} \\ \hline
\multicolumn{1}{|c|}{\textbf{(Baseline)} CQCC+GMM+EM \cite{9todisco2019asvspoof}} & 11.04          & 0.2454 & 9.57          & 0.2366 \\ \hline
\multicolumn{1}{|c|}{\textbf{(Baseline)} LFCC+GMM+EM \cite{9todisco2019asvspoof}} & 13.54          & 0.3017 & 8.09          & 0.2116   \\ \hline
Spect+ResNet+CE \cite{28alzantot2019deep}
& 3.81          & 0.9940& 9.68         & 0.2741 \\ \hline
MFCC+ResNet+CE \cite{28alzantot2019deep}                                                   & -          & - & 9.33         & 0.2042   \\ \hline
Spect+ResNet+CE \cite{18lai2019assert}                                                     & 1.29          & 0.0360 & 11.75          & 0.2160 \\ \hline
Joint-gram+ResNet+CE \cite{10cai2019dku}                                                     & 1.23          & 0.0305 & -          & -  \\ \hline
LFCC+LCNN+A-softmax \cite{30lavrentyeva2019stc}                                                     & 4.60          & 0.1053 & 5.06          & 0.1000 \\ \hline
Spect+LCNN+A-softmax \cite{30lavrentyeva2019stc}                                                     & -         & - & 4.53          & 0.1028   \\ \hline
FG-CQT+LCNN+CE \cite{31wu2020light}                                                     & -          & - & 4.07          & 0.1020 \\ \hline
Spect+LCGRNN+GKDE-softmax \cite{32gomez2020kernel}                                                     & 1.06          & 0.0222 & 3.77          & 0.0842   \\ \hline
Spect+LCGRNN+triplet                                                      & 0.92         & 0.0198 & -          & - \\ \hline
Fbank\&CQT+ResNeWt+CE \cite{33cheng2019replay}                                                     & 0.52 & 0.0134 & - & -   \\ \hline
CQTMGD+ResNeWt+CE \cite{33cheng2019replay}                                                     & 0.94          & 0.0250 & -          & - \\ \hline
Spect+SE-Res2Net50+CE \cite{23li2021replay}                                                     & 0.74          & 0.0207 & 8.73          & 0.2237   \\ \hline
LFCC+SE-Res2Net50+CE \cite{23li2021replay}                                                    & 1.46         & 0.434 & 2.87          & 0.0786 \\ \hline
CQT+SE-Res2Net50+CE \cite{23li2021replay}                                                     & \textbf{0.46}          & \textbf{0.0116} & 2.50          & 0.0743 \\ \hline
Raw signal+SincNet+CE \cite{16zeinali2019detecting}                                                    & -          & - & 20.11          & 0.3563 \\ \hline
logCQT\&powSpect+VGG+CE \cite{16zeinali2019detecting}                                                     & 2.11          & 0.527 & -          & - \\ \hline
LFCC+ResNet18+OCS \cite{24zhang2021one}
                                                     & -          & - & 2.19          & 0.0590 \\ \hline
\textbf{Proposed:} LFCC+SE-ResABNet+CombLoss                                & -          & - & \textbf{1.89}          & \textbf{0.0507}   \\ \hline
\textbf{Proposed:} logPowSpec+EABNet+CombLoss                                & 0.86          & 0.0239 & -          & -   \\ \hline
\end{tabular}
\label{tab:my-table}
\end{table*}
\subsection{Comparison with other single models}
The proposed models have been compared with some of the single models and the baseline models according to the presented objectives. Some of the top-performing models used for relevant purposes are shown and compared with the proposed model in Table 4. For the LA attack, the LFCC+SEResABNet+CombLoss model achieved an EER=1.89\% and t-DCF=0.507, which outperforms the baseline model LFCC-GMM. The proposed model also improved by approximately 0.98\% from its corresponding base model (LFCC+SEResNet50+CE). Also, by comparing the results obtained with other works, it can be seen that this model outperforms LFCC+ResNet18+OCS, which to the best of our knowledge, shows state-of-the-art performance.
For physical access attacks, the LogPowSpec+EFFA0+CombLoss model achieved EER=0.86\% and t-DCF=0.0239. This result is significantly better than the base models. Compared to results reported in the 2019 challenge, the proposed model also appears to outperform 90\% of methods which use fusion models. These results, and other favorable features such as fewer parameters and shorter runtime compared to other models, prove the efficiency of the proposed EABN model.
\section{Conclusion}

Spoof detection is considered a major security concern in authentication systems, particularly the ASV system, demonstrating a clear need for solutions to combat this issue. There are generally two approaches to detecting spoofing attacks on ASV systems: the first is to develop an appropriate classifier targeted specifically at detecting the mentioned attacks, while the second approach is conducted as a preliminary step for extracting discriminative features. In the case of the former, most classifiers fail to consider the issue of optimality in terms of number of parameters and runtime. On the other hand, most proposed models are not interpretable from a human perspective, and features are chosen according to expert's knowledge, and therefore lack generalization to unseen attacks. However, a modular architecture based on branches of attention and perception gives the system the ability to easily utilize any classifier or method to produce an interpretable attention mask and improve classification. To this end, the proposed combined loss function, particularly the triplet center loss, succeeded in yielding a discriminative feature space that can help achieve a more generalized model for unseen attacks.

The proposed model and loss function were evaluated on ASVspoof 2019 data. Using LogPowSpec and LFCC features, along with the first-time use of the EfficientNet-A0 architecture and the well-performed SE-Res2Net50, this study provides a novel method for detecting spoofs. The findings show that, the LFCC+SEResNet50+CE model runs with an EER of 1.89\% and t-DCF of 0.507 in the logical access scenario, which to the best of our knowledge, outperforms all state-of-the-art methods. The EFFA0+CombLoss also obtained an EER of 0.86\% and t-DCF of 0.0239 for the physical access scenario, which is better than 90\% of the models presented for the ASVspoof 2019 challenge. It worth noting that the EfficientNet-A0 consists of only 95,000 parameters. The findings also shed light on certain special cases observed for the produced attention masks. For example, LFCC features outperformed MFCCs in detecting logical access attacks. Alternatively, to detect replay attacks, focusing more on silent segments and frequencies in the human speech frequency range can improve the performance.

\bibliographystyle{IEEEtran}

\bibliography{mybib}

\end{document}